# THE BULGE/DISK CONNECTION IN LATE-TYPE SPIRALS

*Observational Evidence for Models of Secular Evolution*

STÉPHANE COURTEAU
*NOAO/KPNO*
*Kitt Peak National Observatory, 950 N. Cherry Ave., Tucson, AZ 85726*

**Abstract.**
Recent ground-based photometric investigations suggest that central regions of late-type spirals are closely coupled to the inner disk and probably formed via secular evolution. Evidence presented in support of this model includes the predominance of exponential bulges, the correlation of bulge and disk scale lengths, blueness of the bulge and small differences between bulge and central disk colors, detection of spiral structure into the core, and rapid rotation. Recent HST observations show that our own bulge and that of M31, M32, and M33 probably harbor both an old and intermediate-age populations in agreement with models of early collapse of the spheroid plus gas transfer from the disk. Secular evolution provides a mechanism to build-up central regions in late-type spirals; mergers or accretion of small satellites could explain the brighter, kinematically distinct bulges of Sa's and S0's.

## 1. Introduction

In recent years, many studies have addressed the formation and evolution of bulges in spiral galaxies. Models proposed range from the classic picture of dissipational collapse and spheroid formation (Eggen, Lynden-Bell, & Sandage 1962, hereafter ELS; Sandage 1986) to accretion of small satellites (Pfenniger 1991, Zaritsky 1995) and mergers + starbursts (Schweizer 1990). Population studies of extragalactic bulges have been hampered by the difficulty to resolve them into stars. HST observations of our own bulge now provide deeper color-magnitude diagrams and observers have started paying greater attention to properties of bulges, in preparation perhaps for the new era of adaptive optics and high-resolution infrared photometers and spectrographs, both in space and ground-based. Here, I will limit myself to considering only one of the few possible scenarios for the formation of "bulges"[1]. The evidence in favor of secular evolution in late-type spirals is compelling which is why I have chosen to focus on it. This article is based in part on a Letter by Courteau, de Jong, and Broeils (1996). Recommended reading material on the topic of exponential disks, secular evolution, or bar dynamics include Struck-Marcell (1991), Sellwood & Wilkinson (1993), Pfenniger (1993), Martinet (1995), and Pfenniger (1996, this conference). Kormendy (1993) offers an excellent account of observational evidence that "some bulges are really disks".

## 2. An Overview of Secular Evolution

Secular evolution models provide a way to transfer material from the disk into the central regions of a spiral galaxy via angular momentum transfer and redistribution of the initial gas. Models of viscous evolution were first invoked to explain the nature of the exponential distribution of the stars in galactic disks (Silk & Norman 1981, Lin & Pringle 1987, Yoshii & Sommer-Larsen 1989, Saio & Yoshii 1990, Struck-Marcell 1991, hereafter SM91; Olivier et al. 1991). Given comparable timescales for star formation and viscous redistribution of the mass and angular momentum in the disk, one automatically recovers a disk with an exponential luminosity profile. In these models, other timescale combinations would lead to truncated or power-law profiles. Modern N-body realizations of angular momentum transfer which are independent of a viscosity parametrization also yield disk exponential profiles (Pfenniger & Friedli 1991, von Linden et al. 1996). It was also realized that an exponential profile in the central regions is expected from the non-axisymmetric disturbances which will induce inward radial flow of disk material (Hohl 1971, Combes et al. 1990, Saio & Yoshii 1990, Pfenniger & Friedli 1991, SM91, Kormendy 1993, hereafter K93; Pfenniger 1993, hereafter P93).

Efficiency of transport will be improved with a bar or oval distortion which can be triggered by the global dynamical instability of a rotationally supported disk or induced by interactions with a satellite (K93, P93, Martinet 1995, hereafter M95, and references therein). This in turn, will catalyze funneling of disk material into the central regions. Disk material will be heated vertically up to 1–2 kpc above the plane via resonant scattering of stellar orbits by the bar-forming instability. A "bulge-like" component with a nearly exponential profile will emerge due to relaxation induced by the bar. The properties of the disk's central regions are directly coupled to the relative time-scales of star formation and angular momentum transfer. Such a model is expected to produce correlated scale lengths and colors between the disk and its central regions.

---

[1] From this point on, I shall distinguish "bulges" as kinematically and probably chemically distinct entities from the disk, from the "central regions" of late-type disks which will be understood as a central accumulation of disk material. To further clarify this distinction, "central regions" apply to 1-2 kpc radii whereas "core" would be more appropriate for 500 pc or less.

Gas redistribution by the bar can cause its own dissolution. Secular accumulation or satellite accretion of only 1-3% of the total stellar disk mass near the center is sufficient to induce dissolution of the bar into a lens or triaxial component and later into a spheroid (Kormendy 1982, Pfenniger & Norman 1990, P93, Friedli 1994, M95, Norman, Sellwood, & Hasan 1996). It is estimated that about two-thirds of disk galaxies currently have a bar, especially as revealed in the infrared (Block & Wainscoat 1991, Zaritzky, Rix & Rieke 1993, Sellwood & Wilkinson 1993, Martin 1995) and that most spirals have probably harbored a self-destructive bar at one time or another during their evolution (Friedli & Benz 1993, hereafter FB93; Friedli *et al.* 1994). Once a bar has formed, thickened and subsequently dissolved, no more thickening of the central dissipationless material is expected unless triggered by starbust activity or stellar accretion. Secular evolution is a viable mechanism for producing the small, central accumulations of material in late-type disks. Bigger bulges, however, could not be formed this way without disrupting the disk. The energy required to heat up the central material is far greater than the total bar and disk's mechanical energy supply. Accretion of a small satellite to explain the bigger bulges of S0-Sa's provides an appealing alternative (P93).

## 3. Luminosity Profiles

Recently, the groups of Peletier, Balcells, and Andredakis (Andredakis & Sanders 1994, Andredakis, Peletier & Balcells 1995; hereafter APB95), and Courteau, de Jong, Broeils, and Holtzman (de Jong & van der Kruit 1994, de Jong 1995, hereafter dJ95; Courteau, de Jong & Broeils 1996, hereafter CdJB96; Broeils & Courteau 1996, hereafter BC96; Courteau & Holtzman 1996, hereafter CH96) have used their high-quality surface brightness (SB) profiles to show that *central regions of disks are generally best described by an exponential luminosity profile.*

In modeling the stellar density distributions in spiral galaxies, one has often assumed an $r^{1/4}$ brightness law for the central regions (de Vaucouleurs 1948) and an exponential surface density for the outer regions of the disk (de Vaucouleurs 1959, Freeman 1970). Departures from the standard de Vaucouleurs profile in the central light distribution of early and late-type systems are however not new (de Vaucouleurs 1959, van Houten 1961, Frankston & Schild 1976, Kormendy & Bruzual 1978, Burstein 1979, Shaw

& Gilmore 1989, Andredakis & Sanders 1994). For example, Kent, Dame & Fazio (1991) used the Space Shuttle Infrared Telescope to show that the Milky Way bulge is best described by an exponential luminosity profile with a scale length of 500 pc.

A reliable approach to bulge-to-disk (B/D) decompositions is to fit for the shape of the luminosity profile as well. This method was first proposed by Sérsic (1968) with a generalized law of the form

$$\Sigma(r) = \Sigma_\circ exp\{-(r/r_\circ)^{1/n}\}$$

where $\Sigma_\circ$ is the central surface brightness (CSB), $r_\circ$ a scaling radius, and $n$ is the shape parameter. If $n = 1$, one has a pure exponential profile with $r_\circ$ as the scale length; with $n = 4$, one recovers a deVaucouleurs profile. As $n$ becomes large, the Sérsic profile approaches a power law. Caon, Capaccioli, & D'Onofrio (1993) used their data on ellipsoids (E/S0s) and the low surface brightness (LSB) dwarf galaxies of Davies et al. (1988) to show that the parameter $n$ correlates with absolute luminosity and half-light radius, such that bigger, brighter systems have larger values of $n$. This result was extended to brightest cluster galaxies by Graham et al. (1996); see Fig. 1.

Following Caon et al., APB95 and CdJB96 applied the Sérsic law to large samples of spiral galaxies for the first time. APB95 used near-infrared $J$ and $K$ images for 30 early-type spirals and $r$-band data from Kent (1986) for 21 late-type systems and performed 1 dimensional B/D decompositions using the technique of Kent (1985). The shape parameter $n$ is fitted as a free variable and the disk is a fixed exponential. CdJB96 have combined the collection of deep $r$-band profiles of Sb/Sc galaxies by Courteau (1992, 1996; hereafter C96) and $BVRIHK$ photometry of dJ95 for 86 face-on Sa–Sm galaxies. Courteau's Tully-Fisher sample comprises 350 spirals but 243 were kept for final decompositions; the rest had too small a bulge to be resolved successfully. B/D decompositions were done independently by dJ95 and BC96. dJ95 used both major-axis profiles (1D) and full image B/D decompositions[2] of his thesis sample with fixed $n = 1$, 2, and 4 for the central regions and a standard disk exponential. BC96 decompose the light profile of Courteau's galaxies as "bulges" with $n = 1$ and 4 plus an exponential disk, following the method introduced by Kormendy (1977).

"Bulges" of late-type spirals are small and their luminosity profile can be severely affected by atmospheric blur. Both teams used extensive simulations with a wide range of input parameters and various values of $n$ to derive a space of recoverable parameters under specific seeing conditions. Sky subtraction errors, which represent another fundamental source of uncertainty in fitting the SB profile in central regions, were examined carefully. Seeing is accounted for by convolving the model profiles with a Gaussian PSF with a dispersion measured from field stars.

---

[2]2D decompositions offer the advantage of fitting for any additional central component, like a bar, ring or lens. They also yield a more robust recovery of simulated input parameters (Byun & Freeman 1995, dJ95). Still, CdJB96 find that their results do not depend strongly on the type of techniques adopted.

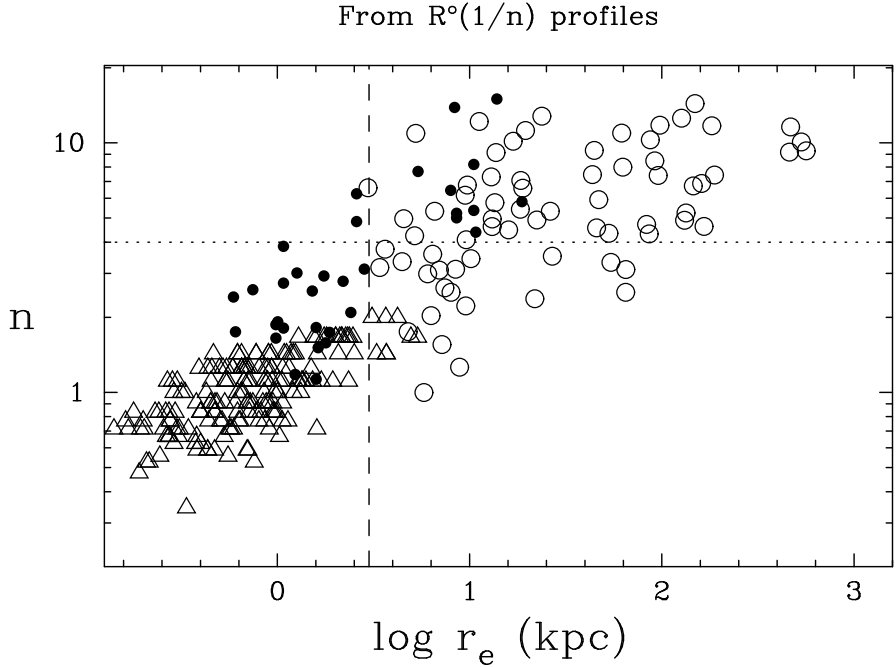

Figure 1. Updated figure for $n$ vs $\log(r_e)$ by Graham et al. (1996) based on Fig. 5 of Caon et al. (1993). The open circles show brightest cluster members from Lauer & Postman (1994), filled circles represent a sample of 33 E/S0 galaxies from Caon et al. (1993), and the triangles are for 187 LSB dwarf galaxies from the study of Davies et al. (1988). Note that these fits cover the entire extent of the luminosity profile and thus, unlike Figure 2, are not confined to the galaxies' central regions. This distribution follows a luminosity trend with LSBs and faint spheroids (E/S0s) at the bottom and giant Es and BGCs at the top. The dashed line at $r_e = 3$ kpc serves to distinguish the fainter and brighter galaxy families. The de Vaucouleurs profile ($n = 4$) is shown with a dotted line; it roughly delineates the hot stellar systems ($n < 4$) from objects that have grown by accretion or mergers ($n > 4$) (Caon et al. 1993).

IR data are extremely useful for studying the central light, unobscured by the dust; profile decompositions are thus less likely to suffer from internal absorption effects (Phillipps et al. 1991). Furthermore, near-IR images are less influenced by star formation or starburst activity near the galaxy center.

APB95 showed that bulge profiles vary with Hubble types and are cor-related with B/D ratios, in good agreement with Davies *et al.* (1988) and Graham *et al.* (1996) for LSBs and spheroids (see Figure 2). CdJB96 provide supporting evidence with their larger sample. From examination of the reduced $\chi^2$ values, they confirmed that most late-type spirals are best fitted by the double-exponential fit. Sa and Sab's are also generally best modeled with a $n = 2$ bulge. Central regions of earlier-type galaxies and of a small fraction ($\sim 15\%$) of the later types are more appropriately fit by a deVaucouleurs law. These results should serve to firmly establish the notion that central regions of late-type spirals are best described by an exponential profile.

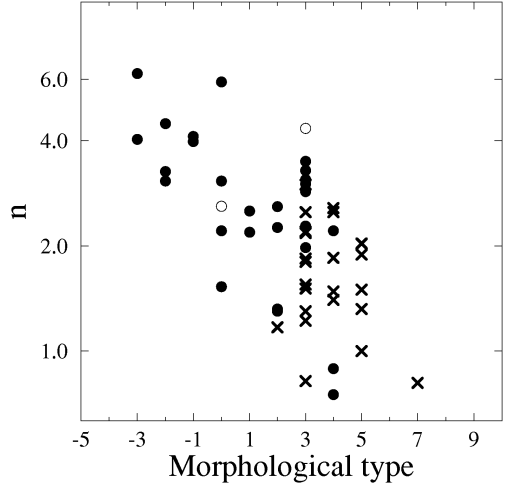

*Figure 2.* Best-fitted parameter $n$ versus morphological type from Andredakis, Peletier, & Balcells (1995; APB). Filled circles consist of 30 S0-Sbc (T=-3 to 4) galaxies with $i > 50°$ from APB. The two open circles are barred galaxies. Crosses represent a small sample of Kent's (1986) photometry of late-type systems. Intrinsically brighter systems have greater $n$; the scatter in this figure is mostly likely correlated with luminosity.

Given that $n = 1$ for most late-type spirals, CdJB96 adopted double-exponential decompositions for all galaxies in their sample to compute a B/D scale length ratio. (Their results is unchanged if restricted to the subsample of pure $n = 1$ central profiles only.) Figure 3 show the measured scale lengths for the joint samples. Combining the two $r$-band data sets, CdJB96 find $r_b/r_d = 0.08 \pm 0.05$ while de Jong galaxies alone at $K'$ yield $r_b/r_d = 0.09 \pm 0.04$ (not shown). Although dust is more conspicuous at $r$ for central regions, $r$-band results statistically reproduce the same range of values than at $K'$, though with wide error bars. Effects of dust are thus not alarming (they become severe at $B$ or $V$).

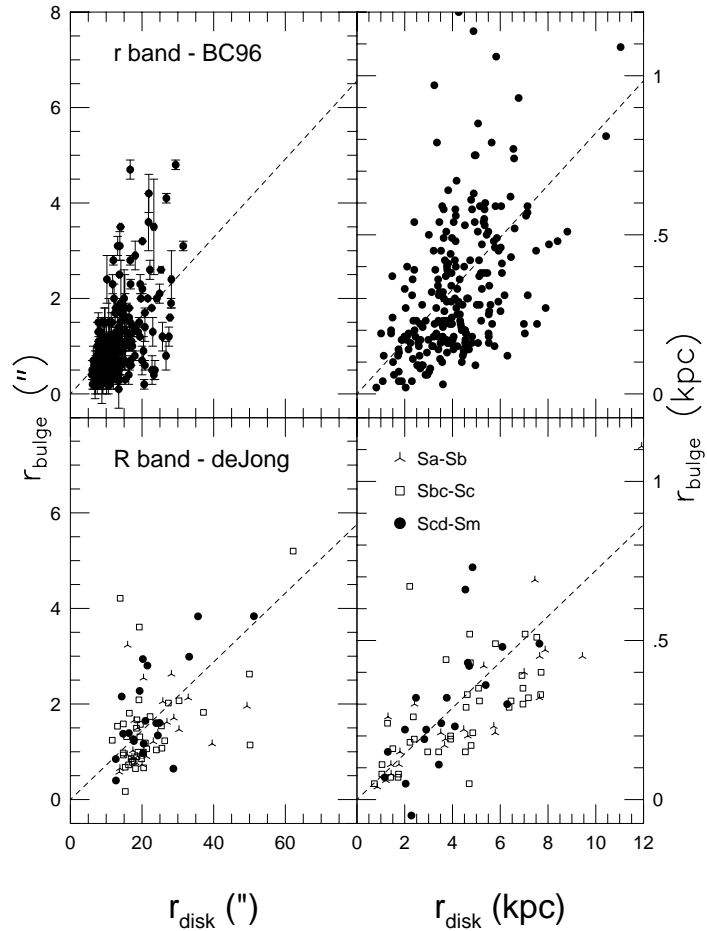

Figure 3. Fitted scale lengths for the bulge and disk from the 1D decomposition of BC96 (top) and the 2D decomposition of de Jong (bottom). Figures on the left are plotted in arcseconds and the ones on the right use absolute coordinates. Apparent units show that the correlation is not affected by resolution effects, and the physical scale allows for a clearer comparison between the two samples. The dashed lines are fits to the data; their slope is, of course, distance independent. All SB profiles were fitted with a double-exponential.

Taken at face value, a correlation of B/D scale lengths is best understood in a model where the disk forms first and the "bulge" that naturally emerges is closely related to the disk. In a scenario where the bulge forms first, it would be hard to understand how a small dynamically hot component could directly influence the disk global structure.

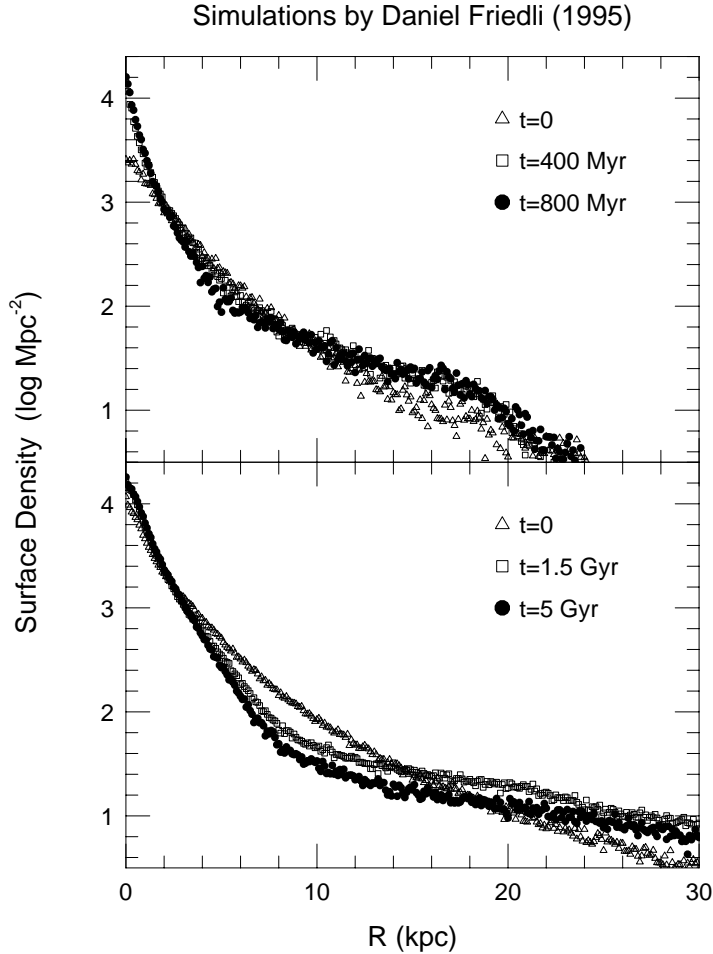

*Figure 4.* The secular evolution of the face-on stellar surface density profile is shown for two independent numerical simulations (top: model $B_{no}$ by Friedli & Benz 1995. Bottom: model by Pfenniger & Friedli 1991). Both simulations originate from an axisymmetric (but bar unstable) Mihamoto-Nagai model and evolve toward an exponential bulge and disk. Though extensive testing of the simulations is still in progress, preliminary checks indicate values for the B/D scalelength ratio close to 0.09. Once the bar has formed, vertical heating of the central regions occurs on short timescales $\lesssim$ 1 Gyr.

*Self-consistent numerical simulations of secular evolution in disk galaxies evolve toward a double-exponential profile with a typical ratio between bulge and disk scale lengths near 0.1* (Friedli 1995, private communication). Two examples of such simulations are shown in Figure 4 (see also Fig. 7 of Norman, Sellwood, & Hasan 1996). It is interesting that some of these

models naturally produce a Freeman (1970) type II disk profile. [3]

Subtraction of an elliptical profile fit from the original galaxy image also shows *residual spiral structure that extends all the way into the center of the galaxy* for the majority of Courteau's thesis sample (Courteau 1992). This provides further support for kinship between the disk and its central regions. (K93, Zaritzky, Rix & Rieke 1993). A bulge could not produce its own spiral structure.

Note that APB95 reject secular evolution on the basis of their continuous spectrum of the index $n$ versus morphological type. They propose that the smooth sequence they observe (Fig. 2) can only result from a single mechanism of bulge formation. Given the large scatter in that diagram, such a conclusion seems ill-based. A scenario in which bright bulges (as in S0 and Sa's) form principally from a minor merger and smaller bulges (Sd's $\rightarrow$ Sab's) form mainly via secular evolution is not likely to leave any bi-modal imprint on the spectrum of "n" as both processes will operate to some degree of efficiency for all Hubble types.

## 4. Color Gradients

To the extent that dust and stellar population effects (age and metallicity) can be disentangled, the successful enterprise of dating disks and their central regions requires optical and IR photometry **and** line-strength gradients[4] (Searle *et al.* 1973, Frogel 1985, Evans 1992, Silva & Elston 1994, Peletier *et al.* 1994, Worthey 1994, de Jong 1995, Just, Fuchs, & Wielen 1996, Peletier & Balcells 1996a,b,c, hereafter collectively as PB96). Accurate photometry of radial profiles or wedge apertures on multicolor images of a disk galaxy indicate that optical color gradients in the disk are relatively small ($\leq 0^{m}.5$) in $B - V$ and $V - R$ but can be as large as 2 mags in $B - H$ or $B - K$ (Elmegreen & Elmegreen 1984, 1985; Courteau & Holtzman 1994, Peletier *et al.* 1994, de Jong 1995). This range is somewhat larger than that reported by Kent (1986) or Terndrup *et al.* (1994). Disk colors also vary greatly among late-spirals.

---

[3]Dust extinction will also conspire to making Type II disk profiles; see Evans (1992), Courteau and Holtzman (1994), Giovanelli *et al.* (1994).

[4]Ideally, one would like to use H$\alpha$ fluxes to measure the current star formation rate, CO 2.36$\mu$m bandhead measurements to constrain the relative contributions of M dwarf, giant, and supergiant light (even in the absence of dust) (Silva 1996), global IRAS fluxes at 60$\mu$m and 100$\mu$m from warm dust heated primarily by O and B stars (Evans 1995, Devereux 1995), FAR-IR fluxes to measure the re-radiated stellar flux absorbed by cold dust, and Balmer spectroscopy without diffuse gas contamination plus the Calcium doublet and the 4000 Å break to lift the age-metallicity degeneracy, . Measurements of the $H - K$ color are necessary to constrain the contribution of intermediate-age AGB stars (Wise & Silva 1996, Silva 1996). Stellar population models which incorporate both age and metallicity effects are necessary.

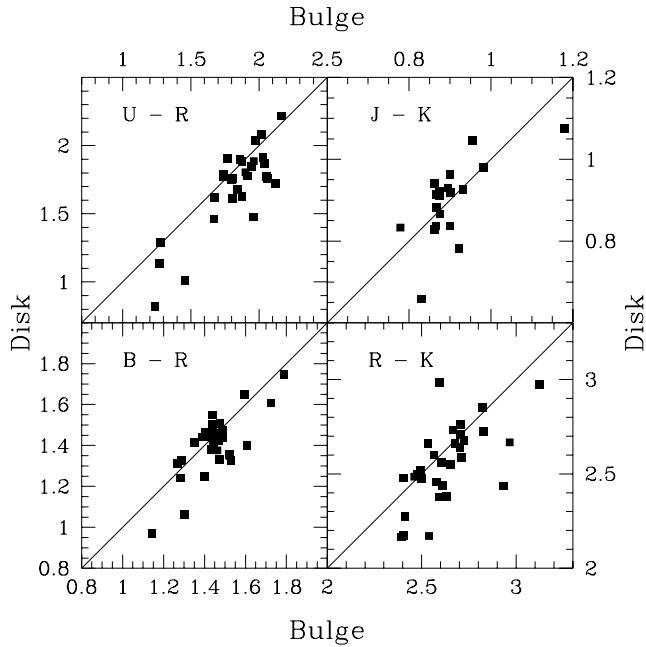

*Figure 5.* Disk colors against bulge colors from Peletier & Balcells (1996). Bulge colors are measured at $r_e/2$ or $5''$, whichever is larger, and disk colors are measured at 2 major-axis scale lengths. Color differences are, on average, of order $0\overset{m}{.}1$ for all passband combinations. Central colors are measured in special regions where extinction is minimal.

Measurement and interpretation of disk *central* colors has essentially neglected the effects of dust reddening prior to the recent work of Balcells and Peletier (1994). Then followed the work of Terndrup *et al.* (1994), dJ95, PB96, and CH96 (see also the article by Frogel *et al.* in these proceedings). The samples of Peletier and Balcells, and dJ95 have been described above. Terndrup *et al.* (1994) observed 43 SO and later-type spirals at $J$ and $K$ with matching $r$-band photometry from Kent (1986) (see also Terndrup 1996). CH96 includes $BVRH$ color gradients for a few hundred Sb-Sc Northern galaxies.

A key question is whether the *stellar population* of the "bulge" is the same as, or related to, that of the disk. Collectively, the authors quoted above have shown, using local colors instead of integrated light, that *central regions of disks are bluer than ellipticals of the same luminosity* and reported *small, negative color gradients. The color difference between the inner disk and its central regions is very small* ($\sim 0\overset{m}{.}1$) *at all colors.* Figure 5 from PB96 summarizes this last statement well. Although PB96 realize that similar colors would support a model of secular evolution, they disregard that possibility on the basis that starbursts would be the only physical process able to build up the observed central space densities in disks. However, secular evolution with bars can manufacture "bulges" with a high degree of rotational support, for late-type systems at least.

PB96 examine the scale length variations in passbands sensitive or "immune" to dust (Evans 1994, Beckman *et al.* 1994) to show that dust effects appear to be small for galaxy types earlier than Sab. Thus for early-type systems, it is assumed that dust effects are negligible and effects of stellar population can be studied directly. Using the population synthesis models of Vazdekis *et al.* (1995), PB96 show that the age difference between the central regions and the inner disks of *early-type spirals* at 2 scale lengths is less than 30 %.

It is worth pointing out that current spectrophotometric population synthesis models for early-types agree poorly in the IR (Charlot, Worthey, & Bressan 1996; hereafter CWB96). These disagreements stem partly form the fact that i) temperatures and numbers of M giants are not well known, and ii) infrared atomic and especially molecular line opacities are not as accurately determined as in the optical. Author-to-author scatter can be large. Scatter in $V - K$ or $R - K$ model colors (*e.g.* see PB96, Table 3) suggests an age uncertainty of 35% even if all other parameters are determined (CWB96). The situation is even worse for younger populations (Charlot 1996). This should not be interpreted as the demise of color gradient studies but current investigations based on such models remain difficult to interpret. The presence of dust in late-type systems makes the analysis of colors + population synthesis models even harder (see e.g. Terndrup 1996). De Jong (dJ95) nonetheless attempted to combine his new 3D radiation transfer code with multiple dust geometries and stellar population models to explain his observed color gradients. His main conclusion, similar to PB96 for early-types, is that dust plays a minor role and the gradients are best explained by a combined stellar age and metallicity gradient across the disk.

In light of the large uncertainties inherent to this sort of population analysis (see CWB96 for details), the safest concluding remark is simply that the similarity between central and inner disk colors, and the fact that bulges are bluer than ellipticals of the same luminosity, are qualitatively consistent with a picture of inward gas transport and subsequent star formation from the inner disk to the center of the galaxy. Age dating of extragalactic bulges will require accurate line-strength gradients and far more robust and complete spectral evolution models (Charlot 1996).

## 5. Line Strengths, Kinematics and Chemical Evolution

Although bulges are known to be bluer than ellipticals of equal absolute luminosity, it has also been known for some time that some "bulges", mostly in barred galaxies, have smaller velocity dispersion than ellipticals of the same $M_B$ (Whitmore, Kirshner, & Schechter 1979, Whitmore & Kirshner 1981, Kormendy & Illingworth 1983, K93). Most of these systems are actually rapid rotators, as shown in the classic $V_{\max}/\sigma - \epsilon$ diagram (Kormendy 1985, K93 and references therein). *Unlike ellipticals, bulges' kinetic energy comes mostly in rotation which must be imparted from the disk*, in agreement with a causal link between the disk and the "bulge".

Chemical evolution is also likely to be affected by gas flows in the center of galaxies. For galaxies undergoing central star formation bursts, Friedli *et al.* 1994 show that the gas-phase abundance gradient should be characterized by two separate slopes corresponding to the inner and outer regions of the disk. A metallicity-velocity dispersion relation for the core is expected as well, though current nuclear stellar abundances are too uncertain to provide conclusive evidence (Friedli & Benz 1995). If metallicity is controlled by the depth of the potential well, low-L disks, which will only create small "bulges", will then have low dispersions and low metallicities (augmented perhaps by self-enrichment during star formation that follows inward transport of disk material). Few studies have addressed the issue of abundances and existence of a fundamental plane for the center of extragalactic disks. Boroson (1980) measured central metallicities ($Mg_2$ index) in the bulge of 24 spirals and found a greater correlation with bulge light than with total light though with poor statistics. A similar correlation and matching of $Mg_2$ with central velocity dispersion by Jablonka *et al.* (1995) also suggest that bulges and ellipticals would occupy the same locus in the fundamental plane (Bender *et al.* 1992, 1993). Such a picture, if true, would suggest a decoupling of the bulge and disk, in stark contrast with models of secular evolution.

Sil'chenko (1993) studied Fe I 5270, Mg I 5175 absorption features in the central regions of many early and late-type spirals and finds that [Mg/Fe] is mostly solar, contrary to the observation of Mg to Fe enrichment in ellipticals, S0 galaxies (Fisher, Franx & Illingworth 1996) and our own Galactic Bulge (McWilliam & Rich 1994). Aided by the Balmer/Ca II age diagnostic, she inferred that disk nuclei contain a significant number of stars less than $5 \times 10^9$ years old. However, Sil'chenko's data set contains only a small number of late-types and her measurements of a mean Balmer equivalent width may be spoiled by diffuse ionized hydrogen emission at H$\beta$. In comparing trends in $Mg_2$-$\sigma$, Jablonka *et al.* are also limited to a small number of late-type spirals (8) and, unlike Sil'chenko, they do not distinguish be-tween age, metallicity and luminosity effects. Note that Fisher, Franx & Illingworth (1996) find comparable bulge and disk ages for S0 galaxies but reject formation scenarios in which bulges formed from heated disk or accreted material at late times. However, their analysis may also suffer, like Sil'chenko's, from the effects of emission contamination at H$\beta$.

At the moment, the status of absorption lines studies for central regions of late-type disks is inconclusive, but suggestive of the co-existence of an old and intermediate-age population. This would be expected, for example, if an old bulge was first formed by dissipational collapse of the primordial gas, and would be later enriched by inward transport of disk material.

Recent HST studies in Baade's window by Ortolani *et al.* (1995) and Rich (1992) and McWilliam & Rich (1994) suggest that *the Milky Way bulge is as old as 47 Tuc and possibly populated by intermediate age stars with some amount of ongoing star formation* (see also Renzini 1996). Minniti *et al.* (1996) show that there is a large spread of metal abundances in the Galactic bulge which could be interpreted as mixing of different components in the inner disk. Finally, Rich (1996, and this conference) interprets the existence of extended giant branches in the bulge of M31, M32, and M33 for a younger population.

## 6. Conclusion

Several simple observational arguments have been presented in favor of models of inward gas transfer for late-type spirals. These are:

- Central surface density best described by an exponential profile
- Restricted range of bulge and disk scalelengths
- Blueness of the central regions and small color difference with the inner disk.
- Rapid rotation and small dispersion for central regions.
- Co-existence of old and younger populations in the center of our Milky Way and other nearby galaxies.

Future tests should include accurate measurement of stellar abundances in extragalactic bulges to unravel the stellar populations and test models of star formation and chemical enrichment (Steinmetz & Müller 1994, Friedli, Benz, & Kennicutt 1994, Friedli and Benz 1995, Mollá & Ferrini 1995, Zaritzky 1995)

## 7. Acknowledgments

I wish to thank my collaborators, Adrick Broeils and Roelof de Jong for their comments on an early draft of this paper and permission to reproduce joint results. I am indebted to Daniel Friedli for sending me various simulations and for valuable comments. Dante Minitti was kind enough to provide many contributions in advance of publication from his ESO proceedings edited with Hans-Walter Rix, "Spiral Galaxies in the Near-Ir". This work also benefited from conversations with Sandra Faber, John Kormendy, Reynier Peletier, Daniel Pfenniger, and Guy Worthey.